\documentclass[11pt,prd,preprintnumbers,eqsecnum,superscriptaddress,nofootinbib]{revtex4}
\usepackage[usenames,dvipsnames]{xcolor}
\usepackage{graphics,epsfig}
\usepackage{graphicx}
\usepackage{color}
\usepackage{amssymb,amsmath,amsfonts}
 \usepackage{amsmath}
\usepackage{amssymb}
\usepackage{hyperref}
\usepackage{epstopdf}

\begin{document}
\title{Solution of the boundary problem for the axial-vector field in the hard-wall  AdS/QCD model}
\maketitle
\begin{center}
      {Nihan Aliyev $^{a,b}$\footnote{e-mail: nihan.aliyev38@gmail.com}  and Shahin Mamedov$^{c,d,e}$\footnote{e-mail : sh.mamedov62@gmail.com}}

\vskip 0.5cm

{\it $^a\,$Faculty of Applied Mathematics and Cybernetics, Baku State University, Z.Khalilov 23, Baku, AZ-1148, Azerbaijan}\\
{\it $^b\,$Institute for Applied Mathematics, Baku State University, Z.Khalilov 23, Baku, AZ-1148, Azerbaijan}\\
{\it $^c\,$Institute for Physical Problems, Baku State University, Z.Khalilov 23, Baku, AZ-1148, Azerbaijan}\\
{\it{$^d$}Institute of Physics, Ministry of Science and Education, H.Javid 33, Baku, AZ 1143, Azerbaijan\\
{\it$^e$}Center for Theoretical Physics, Khazar University, 41 Mehseti Str., Baku, AZ-1096, Azerbaijan}

\end{center}

\thispagestyle{empty}

\vspace{1cm}

\centerline{\bf ABSTRACT} \vskip 4mm

We solve an equation of motion for the axial-vector field under boundary conditions of the bulk-to-boundary propagator in the framework of the hard-wall model of AdS/QCD. The equation is reduced to the form of a homogeneous ordinary differential equation with varying coefficients. We solve conjugate equations and find fundamental solutions. This allows us to establish main relations and necessity conditions. The integral equation has both Volterra and Fredholm terms, and was solved by the iteration method. We apply a new scheme for solving the equation, since all linearly independent necessary conditions for the existence of a solution have been defined and used to establish a sufficient condition for Fredholm’s solvability of the problem.

\vspace{2cm}
\section{Introduction}
The hard-wall model is one of the AdS/QCD models, based on the AdS/CFT correspondence, and successfully describes the phenomenology of strongly interacting elementary particles.  This model is based on conformal symmetry breaking provided by introducing a sharp cutoff in 5-dimensional AdS space at the value $z_m$ from the action for the axial-vector field in the bulk of the 5-dimensional (5D)  Anti de Sitter space \cite{erlich,karch}. A non-normalizable solution to this equation under certain boundary conditions gives a bulk-to-boundary propagator for this field. Over the course of two decades, the question of the solution for the axial-vector field e.o.m. remained unsolved, while the vector and spinor field sectors of the model were solved and studied extensively. For the axial-vector field, as an approximation, the solution for the vector field was applied when the calculation of the axial-vector form factor of the nucleons was performed \cite{mamed1,mamed2,mamed3, mamed4,motoi}. Finding the exact solutions for the axial-vector field's e.o.m. can give an impulse to studies of the physics connected with this field in particle physics. Here, we solve this equation for the axial-vector bulk-to-boundary propagator by combining several methods and tricks to solve different types of differential equations in mathematical physics.

Various direct and inverse problems can be found with ordinary and partial derivatives, problems involving equations having a fractional derivative,  problems with a continuous variable constructing the conjugate equation to ordinary linear differential equations, and finally, problems with continuous multiplicative and powerative derivative equations in the articles provided on the website \cite{nihan}. We list papers and studies from this homepage, that apply methods (or tricks) useful for solving the problem considered and describe the mathematical procedure applied in this paper.  First, the given ordinary, linear, homogeneous differential equation having singularities in coefficients is written in a singularity-free form. A method for constructing the conjugate equation of an ordinary linear differential equation can be found in Ref. \cite{Naym}. The fundamental solution to the expression of the conjugate of the main part of the obtained equation is constructed \cite{Vl}. With the help of this fundamental solution, the integral expressions of the Volterra and Fredholm terms for the search function were obtained from the main relation \cite{Mammadov}. Since the considered equation has varying coefficients, another fundamental solution of the given equation is used to obtain the expression for the derivative of the sought function, and the second basic relation is obtained. The necessary conditions obtained from these two basic relations give all the necessary conditions that are not linearly dependent \cite{nihan,nihan2}. One can also get acquainted with the investigation of the solution of a mixed problem within the framework of nonlocal and global (integral) boundary conditions for a first-order three-dimensional linear hyperbolic type nonhomogeneous equation in Ref.\cite{nihan2}. An investigation of the solution of the boundary value problem under the nonlocal boundary conditions for the two-dimensional Laplace equation can also be seen in \cite{Mustafayeva}. The additional necessary conditions obtained, together with the given boundary condition, allow us to obtain a sufficient condition for the Fredholm nature of the problem posed in \cite{Mustafayeva}. Finally, the resulting integral equation is solved by the successive approximations method, and the resolvent determines the solution. Refs.\cite{hosseini1, hosseini2} were devoted to solving parabolic-type boundary problems. The work \cite{hosseini3} was devoted to solving the  Cauchy problem for the Navier-Stokes equation. In Ref. \cite{jahanshahi}, a solution for the Cauchy-Riemann equation was investigated. In Ref. \cite{jahan 2} the problem for the Navier-Stokes equation was considered. In Ref. \cite{rezapour}, mixed problems for the Navier-Stokes system were considered. In Ref. \cite{rezapour2}, a solution for the problem formulated by Fefferman for the Navier-Stokes equation was considered.
\section{Axial-vector field in the hard-wall model}
  The 5-dimensional (5D) AdS space metric in Poincaré coordinates is
\begin{equation}
ds^2=g_{MN} dx^M dx^N
	= \frac{1}{z^2} \eta_{MN} dx^M dx^N,	\quad \varepsilon<z<z_0.\label{1}
\end{equation}
The $M,N$ indices run over $M,N=0,1,2,3,5$,  $\eta_{MN}=\text{diag}\left(1,-1,-1,-1,-1\right)$. The ultraviolet (UV) boundary of the AdS space is $z= \varepsilon$ with  $\varepsilon \to 0$, and in the AdS/QCD duality, it corresponds to the UV limit of QCD. The infrared (IR) boundary is the cut-off wall located at $z=z_m$. 
 We set the AdS curvature $R=1$.

  The 5D left- and right-handed gauge fields $A_{L} $and  $A_{R }$ corresponding to the chiral flavor symmetry group SU$(2)_L\times$SU$(2)_R$ of the model, and the axial-vector field $A$ composed of these fields:
\begin{equation}
A=\frac{1}{2}\left(A_L-A_R\right). \label{2}
\end{equation}

We will use the gauge fixing choice for the fifth component of the $A$ field $A_z=0$. In addition, in the hard-wall model, the scalar field  $X=X_0\exp(i2\pi^a t^a)$ is introduced, where  $t^a=\sigma^a/2$ and $\sigma^a$ are the Pauli matrices. The expectation value of this field is determined by the classical solution satisfying the UV boundary condition $(2/\epsilon)X(\epsilon) = M$ for the quark mass matrix $M$:
\begin{equation}
  X_0(z) = \frac{1}{2}M
       z + \frac{1}{2} \Sigma  z^3. \label{3}
\end{equation}
The IR boundary condition on $X$ determines the matrix $\Sigma$, and  $\Sigma$ is an input parameter of the model.  As usual, we assume $\Sigma=\sigma \bm1$ and take $M=m_q\bm1$. An exact solution for the $X$ scalar was found in \cite{cherman} and has the following form: 
\begin{equation} \label{4}
    v(z)=\frac{1}{2}am_qz+\frac{1}{2a}\sigma z^{3}, \\
\end{equation}
and $a=\frac{\sqrt{N_c}}{2\pi}$, $m_q$ and $\sigma$ are the real constants. In addition, the solution to the equation for the field $X$ was studied in \cite{ANSME}.
At this stage, the model has four free parameters: $m_q$, $\sigma$, $z_m$, and $g_5$.  The gauge coupling $g_5$ is:
\begin{equation}
g_5^2 = \frac{12 \pi^2}{N_c}\label{5},
\end{equation}
where $N_c=3$ for our choice of gauge symmetry group.

In the axial sector, where we have axial-vector and pseudoscalar fields $A^a_M$ and $\pi$, which describe the $a_1$ and $\pi$ mesons, the action to quadratic order is \cite{erlich}
\begin{equation}
S= \int d^5x\,\left[]
-\frac1{4g_5^2z} F^a_A F^a_A + \frac{v(z)^2}{2z^3} (\partial \pi^a-A^a)^2\right].\label{6}
\end{equation}
In the
$A_z=0$ gauge, separating the transverse and longitudinal modes ($A_\mu= A_{\mu\perp} + \partial_{\mu}\varphi$), the resulting equation of motion obtained from the action (\ref{6}) for the axial vector field $A_{\mu}$ in 4D momentum space has the following form:
\newcommand{\AT}{A^a_\mu}
\begin{equation}
  \left[ \partial_z\left(\frac1z \partial_z \AT \right) + \frac{q^2}z \AT
- \frac{g_5^2 v(z)^2}{z^3} \AT\right]_\perp =0.\label{7}
\end{equation}
 The Fourier components $A_{\mu}(q,z)$ of $A_{\mu\perp}(x,z)$ will be written as $A_{\mu}(q,z)=A_{\mu}(q) A(qz)$. Here, $A(qz)$ is the $z$-dependent part of the wave function, and for the boundary problem considered here, it satisfies the boundary conditions 
 $A_{\mu}(q,z=0)=1$, $ \partial_zA_{\mu}(q,z=z_m)=0$, which are obtained from the boundary terms, when the equation of motion (\ref{7}) is derived from the action (\ref{6}):

\section{The equation} 
Let us solve the boundary problem for $A(qz)$. The equation of motion for $A(qz)$ obtained from (\ref{7}) will be written as follows:
\begin{equation} \label{8}
\left(\frac{A^{\prime}(qz)}{z}\right)^{\prime}+\frac{q^{2}}{z}A(qz)-\frac{g_{5}^{2}}{z^{3}}
v^{2}(z)A(qz)=0, 
\end{equation}
where prime denotes the derivative over the $z$ variable. The boundary conditions at the UV $(z=0)$ and IR $\left(z=z_m\right)$ boundaries are the following: 
\begin{equation} \label{9}
A(0)=1, \ \ \   A^{\prime}(z_{m})=0,
\end{equation}
respectively.

For convenience, let us denote $z^{\prime}=qz$ varying in the region $z^{\prime} \in (0, z_{m}^{\prime}=qz_m=z_0)$. In terms of $z^{\prime}$, equation (\ref{8}) will be written as 
\begin{equation}\label{10}
q^3\partial_{z^{\prime}}\left(\frac{1}{z^{\prime}}\partial_{z^{\prime}}A\left(z^{\prime}\right)\right)+ \frac{q^3}{z^{\prime}}A\left(z^{\prime}\right)-\frac{g_5^2q^3}{z^{\prime3}}\left(\frac{1}{2q}am_qz^{\prime}+\frac{1}{2aq^3}\sigma z^{\prime3}\right)^2A\left(z^{\prime}\right)=0. 
\end{equation}
Having omitted the prime on $z^{\prime}$ in the equation, the derivative term in (\ref{1}) will take the form below:
\begin{equation}\label{11}
\left(\frac{A^{\prime}(z)}{z} \right)^{\prime}=\frac{A^{\prime\prime}(z)z-A^{\prime}(z)}{z^{2}}.
\end{equation}
After dividing by $q^3$ and taking into account (\ref{11}), equation (\ref{10}) will get the form below:
\begin{equation}\label{12}
\frac{A^{\prime\prime}(z)z-A^{\prime}(z)}{z^{2}}+\frac{1}{z}A(z)-\frac{g_{5}^{2}}{z^{3}}z^{2}(\frac{1}{2q}am_q+\frac{1}{2aq^3}\sigma z^{2})^{2}A(z)=0, \ \left(A\left(z_0\right)=0\right).
\end{equation}
Multiplying by $z^2$, the derivative part of this equation can be written in the form below:
\begin{equation} \label{13}
z A^{\prime\prime}(z)-A^{\prime}(z)=z[g_{5}^{2}(\frac{1}{2q}am_q+\frac{1}{2aq^3}\sigma z^{2})^{2}-1]A(z), \ \ \ z \in (0, z_{0}),
\end{equation}
Denoting by
\begin{equation} \label{14}
z[g_{5}^{2}(\frac{1}{2q}am_q+\frac{1}{2aq^3}\sigma z^{2})^{2}-1]=f(z),
\end{equation}
equation (\ref{13}) will be written in a suitable form for solving
\begin{equation} \label{15}
z A^{\prime\prime}(z)-A^{\prime}(z)=f(z)A(z), \ \ \ z \in (0, z_{0}).
\end{equation}
As equation (\ref{15}) is one in the second-order derivative, there are two main relations for it, which we shall establish in the next two sections. Let us first find the conjugate operator for the left-hand side of equation (\ref{15}). 
\section{Conjugate equation}
Using the method, which was described in Ref. \cite{Naym}, it can be easily found that the conjugate to the left-hand side of this equation is of the following form:
\begin{equation} \label{16}
z B^{\prime\prime}(z)+3B^{\prime}(z).
\end{equation}
To this end, let us multiply  the left-hand side of (\ref{15}) by $B(z)$ and integrate by parts:
\begin{eqnarray}
 \int \limits _{0}^{z_{0}}\left[ z A^{\prime\prime}(z)-A^{\prime}(z)\right]B(z)dz= \int \limits _{0}^{z_{0}} z A^{\prime\prime}(z)B(z)dz-\int \limits _{0}^{z_{0}}A^{\prime}(z)B(z)dz=\nonumber \\ \nonumber
=A^{\prime}(z)zB(z)|_{0}^{z_{0}}-\int \limits _{0}^{z_{0}}dz A^{\prime}(z)\left(zB(z)\right)^{\prime}- A(z)B(z)|_{0}^{z_{0}}+\int \limits _{0}^{z_{0}}dz A(z)B^{\prime}(z)= 
\end{eqnarray}
\begin{equation}
=-A(z)\left(zB(z)\right)^{\prime}|_{0}^{z_{0}}+\int \limits _{0}^{z_{0}}dz A(z)\left(zB(z)\right)^{\prime\prime}-A(z_0)B(z_0)+\int \limits _{0}^{z_{0}}dz A(z)B^{\prime}(z)= \nonumber  
\end{equation}
\begin{eqnarray}
=-A(z_0)\left[z_0B^{\prime}(z_0)+B(z_0)\right]+\int \limits _{0}^{z_{0}}dz A(z)\left[B^{\prime\prime}(z)z+2B^{\prime}(z)\right]-A(z_0)B(z_0)+\nonumber \\
+\int \limits _{0}^{z_{0}}dz A(z)B^{\prime}(z)=-A(z_0)\left[B^{\prime}(z_0)z_0+2B(z_0)\right]+\nonumber \\
+\int \limits _{0}^{z_{0}}dz A(z)\left[B^{\prime\prime}(z)z+3B^{\prime}(z)\right].\label{17}
\end{eqnarray}
As seen in the last equality in (\ref{17}), the coefficient of $A(z)$ in the integral term is the operator $B^{\prime\prime}(z)z+3B^{\prime}(z)$. This means that this operator obeys the definition of the conjugate operator.

\section{Fundamental solution to conjugate equation}

Let us construct the fundamental solution to the conjugate equation having the right-hand side of some unknown function $h(z)$:
\begin{equation}\label{18}
zB^{\prime\prime}(z)+3B^{\prime}(z)=h(z).
\end{equation}
 Since equation (\ref{18}) has a solution with varying coefficients, there are two $B(z)$ solutions that do not coincide with each other. These solutions lead to the main relations. 
 Let us find the first main relation
Without the right-hand side, i.e, for the $h(z)=0$ case, equation (\ref{18)} is solved for  $B^{\prime}$ as below:
\begin{eqnarray}
zB^{\prime\prime}(z)+3B^{\prime}(z)=0,\nonumber\\
\frac{dB^{\prime}(z)}{B^{\prime}(z)}=-\frac{3dz}{z}.\label{19}
\end{eqnarray}
For the $B^{\prime}(z)$ derivative, we find the solution below:
\begin{equation}
B^{\prime}(z)=\frac{C_1}{z^3},\label{20}
\end{equation}
where $C_1$ is the integration constant. For the equation having the  $h(z)$  right-hand side, (Eq.(\ref{18})), the $C_1$ constant will be a $z$-dependent function:
\begin{equation}
B^{\prime}(z)=\frac{C_1(z)}{z^3}.\label{21}
\end{equation}
Taking into account this solution for $B^{\prime}$ in (\ref{18}) gives us: 
\begin{equation}
z\frac{C_1^{\prime}(z)z^3-C_1(z)3z^2}{z^6}+3\frac{C_1(z)}{z^3}=h(z). \nonumber \\
\end{equation}
From this one finds the following relation between the functions $C_1^{\prime}(z)$ and $h(z)$:
\begin{equation}
C_1^{\prime}(z)=z^2h(z). \label{22}
\end{equation}
Integrating (\ref{22}), we find the integral relation between $C_1(z)$ and $h(z)$: 
\begin{equation}
  C_1(z)=C_1+\int\limits_0^{z}t^2h(t)dt. \label{23} 
  \end{equation}
From (\ref{21}) $B^{\prime}(z)$ will be written as    
  \begin{equation}
  B^{\prime}(z)= \frac{C_1}{z^3}+ \int\limits_0^z  \frac{t^2}{z^3}h(t)dt.\label{24}
  \end{equation}
 The further integration of $B^{\prime}(z)$ gives  a solution for $B(z)$:
  \begin{eqnarray}
  B(z)=-\frac{C_1}{2z^2}+C_2+\int\limits_0^z d \xi\int\limits_0^{\xi}\frac{t^2}{\xi^3}h(t)dt=C_2+\int\limits_0^z t^2h(t)dt \int\limits_t^z \xi^{-3}d\xi- \frac{C_1}{2z^2}=\nonumber \\
  =C_2-\frac{C_1}{2z^2}-\int\limits_0^zt^2 h(t)\frac{1}{2}\left(z^{-2}-t^{-2} \right)dt  =C_2-\frac{C_1}{2z^2}-\int\limits_0^z\frac{t^2-z^2}{2z^2}h(t)dt.\label{25}
\end{eqnarray}

Finally, applying the method of variation of constants in (\ref{25}), it can be shown that the kernel $B(z,t)$ of $B(z)$ has the form:
\begin{equation} \label{26}
B(z, t)=\theta (z-t)\frac{z^{2}-t^{2}}{2z^{2}}.
\end{equation}
Here $\theta$ is the Heaviside step function:
\begin{equation} \label{27}
\theta(z-t)=\left\{ \begin{array}{c}
1,\ \ \ \ \ \ \ \ \ z>t, \\
\frac{1}{2},\ \ \ \ \ \ \ z=t, \\
0, \ \ \ \ \ \ \ \  z<t
\end{array} \right.
\end{equation}
The first and second order $z$ derivatives of the $B(z,t) $ function are equal to the following ones:
\begin{eqnarray}
B_{z}^{\prime}(z, t)=\theta (z-t)\frac{t^{2}}{z^{3}},\label{28}
\\
B_{zz}^{\prime\prime}(z, t)=\frac{1}{z} \delta(z-t)-3\theta (z-t)\frac{t^{2}}{z^{4}}. \label{29}
\end{eqnarray}
From these derivatives, we can obtain the following second-order differential equation for $B(z,t)$

\begin{equation} \label{30}
z B_{zz}^{\prime\prime}(z, t)+3B_{z}^{\prime}(z, t)=\delta (z-t),
\end{equation}  
where $\delta (z-t)$ is the Dirac $\delta$-function. From equation  (\ref{30}) is seen that  (\ref{26}) is the fundamental solution of this equation.
\section{First main relation}
Using equation (\ref{15}) multiplied by (\ref{26}) and integrating over $z$ in the interval $(0,z_0)$,  we shall obtain the following:
\begin{equation}\label{31}
\int \limits _{0}^{z_{0}} z A^{\prime\prime}(z)B(z, t)dz-\int \limits _{0}^{z_{0}} A^{\prime}(z)B(z, t)dz=
\int \limits _{0}^{z_{0}} f(z)A(z)B(z, t)dz.
\end{equation}
Integrating by parts the left-hand side and taking into account (\ref{30}), we get the first main relation as follows:

\begin{equation}
\int \limits _{0}^{z_{0}} f(z)A(z)\theta(z-t)\frac{z^{2}-t^{2}}{2z^{2}}dz+A(z_{0})\theta(z_{0}-t)
\frac{z_{0}^{2}-t^{2}}{z_{0}^{2}}+z_{0}A(z_{0})\theta(z_{0}-t)\frac{t^{2}}{z_{0}^{3}}= \nonumber 
\end{equation}

\begin{eqnarray} \label{32}
    =\int^{z_0}_0 \delta(z-t) A(z)dz=\begin{cases}
        A(t), &\text{$t\in (0, z_{0})$}\\
        \frac{1}{2}A(t), &\text{$t=0$, $t=z_{0}$} .\\
    \end{cases}
\end{eqnarray}

This is the first main relation, and it consists of two parts. The first part is the arbitrary solution of equation (\ref{15}) in $t\in(0,z_0)$.
\begin{equation} \label{33}
A(z_{0})=\frac{1}{2}-\frac{1}{2}\int\limits_{0}^{z_{0}}f(z)A(z)dz,
\end{equation}
Here we have used the boundary condition (\ref{12}) and the definition (\ref{27}) ($B(-t)=0$ since $t>0$). From equation (\ref{32}),  two necessary conditions are obtained, one of which is
\begin{equation} \label{34}
A(t)=\int\limits_{t}^{z_{0}}f(z)A(z)\frac{z^{2}-t^{2}}{2z^{2}}dz+A(z_{0}),
\end{equation}
and the second one turns to identity.
From the solution (\ref{33}), it is seen that the equation obtained for  $A(z)$ consists of both Volterra and Fredholm terms.
\section{Second conjugate operator and fundamental solution}
Let us multiply equation (\ref{15}) by $B^{\prime}(z)$ and integrate in the interval $\left(0, z_0\right)$:
\begin{equation} \label{67}
\int_0^{z_0}z A^{\prime\prime}(z)B^{\prime}(z)dz-\int_0^{z_0}A^{\prime}(z)B^{\prime}dz(z)=\int_0^{z_0}f(z)A(z)B^{\prime}(z)dz.
\end{equation}
Integrating by parts the first integral, we obtain:
\begin{equation} \label{68}
-\int_0^{z_0}A^{\prime}(z)\left[zB^{\prime\prime}(z)+2B^{\prime}(z)\right]dz=\int_0^{z_0}f(z)A(z)B^{\prime}(z)dz.
\end{equation}
Here we have adopted $A^{\prime}(0)=0$, the rightness of which we shall see from the necessity condition (\ref{38})-(\ref{37}).

Let us find the fundamental solution of the following equation:
\begin{equation}
zB^{\prime\prime}(z)+2B^{\prime}(z)=h(z).\label{69}
\end{equation}
\begin{equation}
zB^{\prime\prime}(z)+2B^{\prime}(z)=0.\label{70}
\end{equation}
\begin{equation}
zB^{\prime\prime}(z)=-2B^{\prime}(z).
\end{equation}
\begin{equation}
\frac{dB^{\prime}(z)}{B^{\prime}(z)}=-2\frac{dz}{z}.
\end{equation}
Solving this equation, we find the solution $B(z)$ of the homogeneous equation (\ref{70}) :
\begin{equation}
B(z)=C_1z^{-1}+C_2,
\end{equation}
where $C_1$ and $C_2$ are the integration constants.

Now, we find a general solution of the non-homogeneous equation (\ref{69}) using the variation of constants method. $B(z)$ will be taken
\begin{equation}
B(z)=\frac{C_1(z)}{z}+C_2(z),
\end{equation}
and its derivative will be as:
\begin{equation}
B^{\prime}(z)=\frac{C_1^{\prime}(z)z-C_1(z)}{z^2}+C_2^{\prime}(z).\label{75}
\end{equation}
Adopting 
\begin{equation}
-\frac{C_1^{\prime}(z)}{z}+C_2^{\prime}(z)=0,\label{76}
\end{equation}
from (\ref{75}) we find $B^{\prime}(z)$:
\begin{equation}
B^{\prime}(z)=\frac{C_1(z)}{z^2}.\label{77}
\end{equation}
Taking a derivative from this $B^{\prime}(z)$, and taking into account the obtained $B^{\prime\prime}(z)$ and (\ref{77}) in (\ref{69}) gives us the equation relating $C_1^{\prime}(z)$ and $h(z)$:
\begin{eqnarray}
z\frac{C_1^{\prime}(z)z^2-C_1(z)2z}{z^4}+2\frac{C_1(z)}{z^2}=h(z),\nonumber\\
C_1^{\prime}(z)=zh(z).\label{78}
\end{eqnarray} 
A solution to this equation has the form:
\begin{equation}
C_1(z)=C_1+\int_0^zth(t)dt.\label{79}
\end{equation}
If we take into account (\ref{79}) in (\ref{76}) one also finds $C_2(z)$:
\begin{eqnarray}
-h(z)+C_2^{\prime}(z)=0, \nonumber \\
C_2(z)=C_2+\int_0^zh(t)dt.\label{80}
\end{eqnarray}
Finally, $B(z)$ has the following explicit form:
\begin{equation}
    B(z)=-\frac{C_1}{z}-\frac{1}{z}\int_0^tth(t)dt+C_2+\int_0^zh(t)dt=C_2-\frac{C_1}{z}+\int_0^z\left(1-\frac{t}{z}\right)h(t)dt.\label{81}
\end{equation}
The solution (\ref{81}) is the general solution of equation (\ref{69}), so $C_2-\frac{C_1}{z}$ is the general solution of the equation homogeneous to (\ref{69}), and $\int_0^z\left(1-\frac{t}{z}\right)h(t)dt$ is the one solution of a nonhomogeneous equation (\ref{69}). So, the fundamental solution of equation (\ref{69}) has the following form:
\begin{equation}
B(z,t)=\theta (z-t)\left(1-\frac{t}{z}\right). \label{82}
\end{equation}
Taking first and second derivatives from (\ref{82}) 
\begin{eqnarray}
B^{\prime}_z(z,t)=\delta(z-t)\left(1-\frac{t}{z}\right)+\theta (z-t)\frac{t}{z^2}=\theta (z-t)\frac{t}{z^2}, \label{83} \\
B^{\prime\prime}_{zz}(z,t)=\delta(z-t)\frac{t}{z^2}-2\theta (z-t)\frac{t}{z^3}, \label{84}
\end{eqnarray}
and by putting them on the left-hand side of equation (\ref{69})
\begin{equation}
z\left[\delta(z-t)\frac{t}{z^2}-2\theta (z-t)\frac{t}{z^3}\right] +2\theta (z-t)\frac{t}{z^2}=\delta(z-t), \label{85}
\end{equation}
it can be easily checked that solution (\ref{82}) is the fundamental solution of (\ref{69}):
\begin{equation}
zB^{\prime\prime}_{zz}(z,t) +2B^{\prime}_{z}(z,t)=\delta(z-t). \label{86}
\end{equation}
\section{Second main relation and the necessity conditions}
To obtain the second main relation for equation (\ref{15}), we take the second $B(z,t)$ found in (\ref{82}).
Multiplying (\ref{15}) by $B_z^{\prime} (z,t)$ (\ref{83}) and integrating by parts, we obtain the following second main relation:
\begin{equation}
 -\int \limits _{0}^{z_{0}} f(z)A(z)\theta(z-t)\frac{t}{z^{2}}dz=\int\limits_{0}^{z}\delta(z-t)A^{\prime}(z)dz= \nonumber 
\end{equation}
\begin{equation}
=\left\{ \begin{array}{c}
A^{\prime}(t),\ \ \ \ \ \  t\in (0, z_{0}),\\
\\
\frac{1}{2}A^{\prime}(t),\ \ \ \ \ \ \ t=0, \ \ \ t=z_{0}. \label{37}
\end{array} \right.
\end{equation}

The integral (\ref{37}) also consists of two parts. The second part of the right-hand side of (\ref{37}) $(t=0, t=z_0)$ gives two necessary conditions, since the left-hand side of this relation becomes zero at these values. From (\ref{37}), we get the following two conditions:
\begin{equation} \label{38}
A^{\prime}(0)=A^{\prime}(z_{0})=0.
\end{equation}
It can be seen from (\ref{16}) that the expression obtained from (\ref{16}) corresponds to the expression from (\ref{11}), and there is no contradiction between them. Taking into account (\ref{12}) in (\ref{13}) we shall obtain 
\begin{equation} \label{39}
A(t)=\int\limits^{z_{0}}_{t}f(z)\frac{z^{2}-t^{2}}{2z^{2}}A(z)dz-\frac{1}{2}\int\limits_{0}^{z_{0}}f(z)A(z)dz+\frac{1}{2}.
\end{equation}
As seen from the right-hand side of (\ref{37}) $z>t$. Equation (\ref{39}) is the integral equation having the polynomial kernel. In this equation, the first term is the Volterra term and the second is the Fredholm term. The equations containing both Volterra and Fredholm type terms were studied in \cite{Mammadov}.
\section{Iterations in the Volterra term}
Let us make an iteration in the Volterra term. To this end, denote 
\begin{equation} \label{40}
\frac{t^{2}-z^{2}}{2z^{2}}f(z)=F(t, z).
\end{equation}
Then equation (\ref{39}) accepts the following form:
\begin{equation} \label{41}
A(t)=\int\limits_{z_{0}}^{t}F(t,z)A(z)dz-\frac{1}{2
}\int\limits_{0}^{z_{0}}f(z)A(z)dz+\frac{1}{2}.
\end{equation}

\begin{equation}\label{42}
A(z)=\int\limits_{z_{0}}^{z}F(z,\xi)A(\xi)d\xi-\frac{1}{2}\int\limits_{0}^{z_{0}}f(\xi)A(\xi)d\xi+\frac{1}{2}.
\end{equation}
Writing these expressions for $A(z)$ and $A(t)$ in the Volterra term of the equation (\ref{41}) let us make the first iteration:
\begin{eqnarray}\label{43}
A(t)=\int\limits_{z_{0}}^{t}F(t,z)dz \left[\int\limits_{z_{0}}^{z}F(z,\xi)A(\xi)d\xi
-\frac{1}{2
}\int\limits_{0}^{z_{0}}f(\xi)A(\xi)d\xi+\frac{1}{2} \right]- \nonumber
\\
-\frac{1}{2}\int\limits_{0}^{z_{0}}f(z)A(z)dz +\frac{1}{2}=
\int\limits_{z_{0}}^{t}A(\xi)d\xi \int\limits_{\xi}^{t}F(t,z)F(z, \xi)dz+ \nonumber
\\
+\frac{1}{2} \int\limits_{z_{0}}^{t}F(t,z)dz\left[1-\int\limits_0^{z_0}f(\xi)A(\xi)d\xi \right]
+\frac{1}{2}\left[1-\int\limits_0^{z_{0}}f(\xi)A(\xi)d\xi\right],
\end{eqnarray} 
Denote
\begin{equation} \label{44}
F(t,z)=F_1(t,z),
\end{equation}
 We obtain for the iteration of this kernel
\begin{equation} \label{45}
\int\limits_{\xi}^{t}F_{1}(t,z)F_{1}(z,\xi)dz=F_{2}(t,\xi).
\end{equation}
Then (\ref{43}) becomes
\begin{equation} \label{46}
A(t)=\int\limits_{z_{0}}^{t}F_{2}(t,\xi)A(\xi)d\xi+\frac{1}{2}\left[1-\int\limits_{0}^{z_{0}}
f(\xi)A(\xi)d\xi\right] \left[1+\int\limits_{z_{0}}^{t}F_{1}(t,z)dz\right].
\end{equation}
Let us make one more iteration in (\ref{46}). To this end, we have to find $A(\xi)$ from the equation (\ref{41}) and take it into account in the Volterra term of (\ref{46})
\begin{equation}
A(\xi)=\int\limits_{z_{0}}^{\xi}F_{1}(\xi,z)A(z)dz+\frac{1}{2}\left[1-\int\limits_{0}^{z_{0}}
f(z)A(z)dz\right], \label{47}
\end{equation}
\begin{eqnarray}
A(t)=\int\limits_{z_{0}}^{t}F_{2}(t,\xi)d\xi \left\{ \int\limits_{z_{0}}^{\xi}F_{1}(\xi,z)A(z)dz+\frac{1}{2}\left[1-\int\limits_{0}^{z_{0}}
f(\eta)A(\eta)d\eta\right]\right\}+ \nonumber
\\
+\frac{1}{2}\left[1-\int\limits_{0}^{z_{0}}f(\xi)A(\xi)d\xi\right] \left[1+\int\limits_{z_{0}}^{t}F_{1}(t,z))dz\right]= \nonumber
\\
=\int\limits_{z_{0}}^{t}A(z)dz\int\limits_{z}^{t}F_{2}(t,\xi)F_{1}(\xi,z)d\xi+\frac{1}{2}\left[1-\int\limits_{0}^{z_{0}}
f(\eta)A(\eta)d\eta\right]\left[1+\int\limits_{z_{0}}^t\left[F_1(t, z)+F_2(t,z) \right]dz \right].\label{48}
\end{eqnarray}

Here
\begin{equation} \label{49}
\int\limits_{z}^{t}
F_{2}(t,\xi)F_{1}(\xi, z)d\xi=
F_{3}(t, z),
\end{equation}
is the second iteration of the $F_1(z,t)$ kernel. Then from (\ref{48}) we obtain the following:
\begin{equation} 
A(t)=\int\limits_{z_{0}}^{t}F_{3}(t,z)A(z)dz
+\frac{1}{2}\left[1-\int\limits_{0}^{z_{0}}
f(\eta)A(\eta)d\eta\right]\left[1+\int\limits_{z_{0}}^{t}\left[F_{1}(t, z)+F_{2}(t,z) \right]dz\right].\label{50} 
\end{equation}
For the $n$-th iteration, we can write:
\begin{equation}
A(t)=\int\limits_{z_{0}}^{t}F_{n+1}(t,z)A(z)dz
+\frac{1}{2}\left[1-\int\limits_{0}^{z_{0}}
f(\eta)A(\eta)d\eta\right]\left[1+\int\limits_{z_{0}}^{t}\sum_{k=1}^nF_{k}(t, z) dz\right].\label{51}
\end{equation}
Now we can write $n$-th iteration of the equation (\ref{41})
\begin{equation} 
\begin{array}{cc}
A(t)=\int\limits_{z{0}}^t F_{n+1}(t,z)A(z)dz-\frac{1}{2}\int\limits_{0}^{z_{0}}
f(\xi)\int\limits_{z_{0}}^{t}\left[\sum\limits_{k=1}^{n}F_{k}(t, z)\right]dzA(\xi)d\xi  \\ 
\\
+\frac{1}{2}\int\limits_{z_0}^t\left[\sum\limits_{k=1}^{n}F_{k}(t, z)\right]dz+\frac{1}{2}\left[1-\int\limits_0^{z_{0}}f(\eta)A(\eta)d\eta\right].
\end{array}\label{52}
\end{equation}
If the kernel (\ref{40}) of the Volterra term in the integral equation (\ref{39}) is restricted, i.e., the kernel $F_1(z,t)$ is smaller than some constant $\mathcal{F}$ (\ref{44})
\begin{equation} \label{53}
|F(t,z)|=|F_{1}(t,z)|\leq \mathcal{F},
\end{equation}
$\mathcal{F}$ is a real unknown number. An estimate of its value is given in Appendix A. 
Then from (\ref{45}) we get
\begin{equation}
|F_{2}(t,\xi)|=\left|\int\limits_{\xi}^{t}F_{1}(t,z)F_{1}(z,\xi)dz \right |\leq 
\int\limits_{t}^{\xi}\mathcal{F}\mathcal{F}dz=\mathcal{F}^{2}(\xi-t). \label{54}
\end{equation}
As seen in (\ref{43}) $\xi>t$. From (\ref{49}) we get the following:
\begin{equation}
|F_{3}(t,z)|=\left|\int\limits_{z}^{t}F_{2}(t,\xi)F_{1}(\xi,z)d\xi \right |\leq 
\int\limits_{t}^{z}\mathcal{F}\mathcal{F}^{2}(\xi-t)d\xi=\mathcal{F}^{3}\frac{(z-t)^{2}}{2!}.\label{55}
\end{equation}
Then we obtain a formula for the $n$-th iteration of the $F$ function
\begin{equation} \label{56}
|F_{n+1}(t,z)|\leq \mathcal{F}^{n+1}\frac{(z-t)^{n}}{n!},
\end{equation}
It is seen from (\ref{56}) in the limit $n\to\infty$ $F_{n+1}(t,z)\to 0$ in (\ref{52}) that one obtains:
\begin{equation} \label{57}
A(t)=\frac{1}{2}\left[1-\int\limits_{0}^{z_{0}}
f(\eta)A(\eta)d\eta\right]K(t).
\end{equation}
Thus, the equation becomes a Fredholm equation of the second kind. Here 
\begin{equation} \label{58}
K(t)=1+\int\limits_{z_{0}}^{t}
\sum\limits_{k=1}^{\infty}F_{k}(t,z)dz
\end{equation}
is the kernel of the equation.
\section{Solution to the equation with the 
Fredholm term and degenerated kernel}
The integral equation (\ref{58}) is a second-kind Fredholm integral equation with a degenerate kernel. The solution of the integral equation is the following.
\begin{equation}
A(t)=-\frac{1}{2}K(t)\int\limits_{0}^{z_{0}}
f(\xi)A(\xi)d\xi+\frac{1}{2}K(t), \label{59}
\end{equation}
\begin{equation}\label{60}
\int\limits_{0}^{z_{0}}f(\xi)A(\xi)d\xi=C,
\end{equation}
Here, $C$ is the constant, but it depends on an unknown function $K(t)$

\begin{equation} \label{61}
A(t)=\frac{1}{2}(1-C)K(t),
\end{equation}
This means that
\begin{eqnarray} 
C=\frac{1}{2}\int\limits_{0}^{z_{0}}f(\xi)(1-C)K(\xi)d\xi =\nonumber \\=\frac{1}{2}\int\limits_{0}^{z_{0}}f(\xi)K(\xi)d\xi - \frac{1}{2}C\int\limits_{0}^{z_{0}}f(\xi)K(\xi)d\xi\label{62}
\end{eqnarray}
From (\ref{62}) we get
\begin{equation}
C+\frac{1}{2}C\int\limits_{0}^{z_{0}}f(\xi)K(\xi)d\xi=\frac{1}{2}\int\limits_{0}^{z_{0}}f(\xi)K(\xi)d\xi.\label{63}
\end{equation}
If we take into account (\ref{60}) in (\ref{61}) divide it by
\begin{equation} \label{64}
1+\frac{1}{2}\int\limits_{0}^{z_{0}}f(\xi)K(\xi)d\xi \neq 0,
\end{equation}
we find an expression for the constant $C$:
\begin{equation} \label{65}
C=\frac{\int\limits_{0}^{z_{0}}f(\xi)K(\xi)d\xi}{2+\int\limits_{0}^{z_{0}}f(\xi)K(\xi)d\xi}. 
\end{equation}
Taking into account (\ref{65}) in (\ref{61}) we obtain
\begin{equation} \label{66}
A(t)=\frac{1}{2}\left[1-\frac{\int\limits_{0}^{z_{0}}f(\xi)K(\xi)d\xi}{2+\int\limits_{0}^{z_{0}}f(\xi)K(\xi)d\xi}\right]K(t)=\frac{K(t)}{2+\int\limits_{0}^{z_{0}}f(\xi)K(\xi)d\xi}.
\end{equation}
Thus, we find the solution to equation (\ref{8})-(\ref{9}) in the form (\ref{66}).

{\bf Theorem:} If $q$,$g_5$,$m_q$ and $\sigma$ are given real constants, $z_0>0$, $v(z)$ is polinomial in the form of (\ref{4}) and $f(z)$ is a function in the form of (\ref{14} ), if conditions (\ref{53}) and (\ref{64}) satisfy, then equation (\ref{8})-(\ref{9}) has a solution and this solution is in the form of (\ref{66}).

\section{Appendix}

The value of the constant $\mathcal{F}$  is evaluated from the following condition:
\begin{equation}
    \frac{z^2-t^2}{2z^2}f(z)<\mathcal{F} \ \ (z>t)
\end{equation}
 Explicitly, this condition will be written as follows:
\begin{equation}\label{68}
\left|z\frac{t^2-z^2}{2z^2}\left[g_5^2\left(\frac{1}{2q}am_q+\frac{1}{2aq^3}\sigma z^2\right)^2-1\right]\right|<\mathcal{F}
\end{equation}
We have two cases for estimating $\mathcal{F}$ from this non-equality. 
 
{\it First case}: the second term is larger than the first term in the square bracket in (\ref{68}). In this case, we can omit the modulus and throw away the first term:
\begin{equation}
    z\frac{z^2-t^2}{2z^2}<\mathcal{F}
\end{equation}
Since $z>t$, let us ignore $t^2$ in this non-equality:
\begin{equation}
\frac{z}{2}<\mathcal{F}.
\end{equation}
So $\mathcal{F}$ can be
\begin{equation}
\mathcal{F}=\frac{z_0}{2}.
\end{equation}

{\it Second case}: the modulus of the first term is larger than the modulus of the second term in the square bracket. In this case, we throw away the second term in this bracket. In addition, we ignore $t^2$, since $z>t$. Finally, we get: 
\begin{equation}
\frac{z}{2}g_5^2\left(\frac{1}{2q}am_q+\frac{1}{2aq^3}\sigma z^2\right)^2. 
\end{equation}
Then, the estimated value of $\mathcal{F}$ will be equal to the largest value of this expression:
\begin{equation}
 \frac{z_0}{2}g_5^2\left(\frac{1}{2q}am_q+\frac{1}{2aq^3}\sigma z_0^2\right)^2=\mathcal{F}
\end{equation}

\end{document}